\documentclass[12pt]{iopart} 

\usepackage{iopams}
\usepackage{graphicx}
\usepackage{cite}
\usepackage{float}
\usepackage{color}

\begin{document}

\title[Correlations in Heusler compounds]
      {Electron correlations  in Co$_2$Mn$_{1-x}$Fe$_x$Si Heusler
       compounds.}

\author{Stanislav Chadov, Gerhard~H.~Fecher, and Claudia~Felser}
\ead{chadov@uni-mainz.de}
\address{Institut   f\"ur  Anorganische  und   Analytische  Chemie,
Johannes Gutenberg - Universtit\"at Mainz,  55099 Mainz,
Germany}

\author{Jan~Min\'{a}r, J{\"u}rgen~Braun, and Hubert~Ebert}

\address{Dept.~Chemie   und    Biochemie,   Physikalische   Chemie,
Universit\"at M\"unchen, Butenandtstr.  5-13, 81377 M\"unchen, Germany}

\begin{abstract}

This study presents the effect of local electronic correlations on the
Heusler compounds Co$_2$Mn$_{1-x}$Fe$_x$Si as a function of the
concentration $x$. The analysis has been performed by means of
first-principles band-structure calculations based on the local
approximation to spin-density functional theory (LSDA). Correlation
effects are treated in terms of the Dynamical Mean-Field Theory (DMFT)
and the LSDA+U approach. The formalism is implemented within the
Korringa-Kohn-Rostoker (KKR) Green's function method.

In good agreement with the available experimental data the magnetic and
spectroscopic properties of the compound are explained in terms of
strong electronic correlations. In addition the correlation effects
have been analysed separately with respect to their static or dynamical
origin. To achieve a quantitative description of the electronic
structure of Co$_2$Mn$_{1-x}$Fe$_x$Si both static and dynamic
correlations must be treated on equal footing.
\end{abstract}


\vspace{2pc}

\submitto{\JPD}

\maketitle

\section{Introduction}

One of the most interesting class of materials for magneto-electronic
applications is found in the family of Co$_2$YZ half-metallic Heusler
ferromagnets~\cite{KWS83,FFB07}. Special technological attention has
been paid to the Co$_2$MnSi compound because of a predicted large
minority band gap (0.4~eV) and a very high Curie temperature (985~K)
among the variety of Heusler
compounds~\cite{FSIA90,BNWZ00,FKW+06,KFF07b}. However, an even more
interesting material, Co$_2$FeSi, was found in recent investigations
\cite{WFK+05,WFK+06,KFF06} exhibiting a Curie temperature of about
1100~K and a very high magnetic moment of 6~$\mu_{\rm B}$ in the
primitive cell. Later on it was shown that the complete substitutional
series Co$_2$Mn$_{1-x}$Fe$_x$Si crystallises in the Heusler type
$L2_{1}$ structure with a high degree of structural
order~\cite{BFK+06}. Also confirmed was a Slater-Pauling behaviour
(see~\cite{Kub00,GDP02,FKW+06}) of the magnetic moment ranging linearly
from 5~$\mu_{\rm B}$ to 6~$\mu_{\rm B}$ with increasing Fe
concentration $x$.

A detailed understanding of the electronic and magnetic structure of
half-metallic ferromagnets, their surfaces and corresponding
nanostructures is directly connected with an improved theoretical
description based on first-principle methods. From previous studies
\cite{KFFS06,FBO+07} it follows that quantitative calculations of the
electronic structure of Co$_2$Mn$_{1-x}$Fe$_x$Si compounds require a
correct incorporation of local correlation effects. For example, a
former study \cite{KFFS06} applying the LSDA+U method \cite{AZA91,CS94}
was able to reproduce the experimental band gap as well as the correct
values of the magnetic moments when using reasonable values for the
parameters $U$ and $J$ and when also including the appropriate
double-counting corrections. Similar results \cite{BFK+06} were
obtained by applying the GGA approach which also takes into account
correlations beyond the LSDA-scheme. In the present analysis electronic
correlation effects have been treated in the framework of the Dynamical
Mean-Field Theory \cite{KSH+06}, which quantitatively takes into
account dynamical correlation effects, in particular spin-flip
processes induced by fluctuations. In combination with the LSDA
(LSDA+DMFT) this formalism provides a very reliable approach to deal
with a wide range of static and spectroscopic properties of 3$d$
transition metals. These are, for example, total energies,
magnetic moments, magneto-optical properties and photoemission
intensities \cite{LKK01,GDK+06,PCE03,MEN+05,BME+06,CMK+08}.

The most serious complication when combining LSDA with DMFT arises from
the double-counting problem. As it has been indicated by numerous DMFT
studies, the static many-body effects which are typically over counted
in LSDA+DMFT calculations are relatively small in 3$d$ transition
metals. Therefore, 
the established procedure concerning the description of spectroscopic
properties is to neglect the static part of the self-energy, setting
this term to zero at the Fermi level \cite{LKK01,GDK+06,KL02}. However,
it was recently shown \cite{BME+06}, that this approximation is not
sufficient for an accurate description of angle-resolved photoemission
spectra of Ni. In this case an additional static polarisation term,
which is spin- and orbital-dependent had to be included in the
calculations. Later it was demonstrated \cite{CMK+08} that using LSDA+U
as a static limit for DMFT leads to an improved description of the
orbital and spin magnetic moments in a wide range of the
3$d$ transition metal compounds.

Another recent investigation on Co$_2$MnSi~\cite{CSA+08}, which is
based on the assumption that static correlations are already accounted
by the LSDA, thus considering only dynamical contributions in the 
self-energy, supports the existence of the so-called non-quasiparticle
states~\cite{IKL07}. However, the insufficient treatment of static
correlations resulted in a wrong position of the Fermi level, which in
the case of Co$_2$MnSi~\cite{KFFS06,FBO+07} must be situated closer to
the lower edge of the band gap. Similar findings can be obtained from
former investigations~\cite{IFKA95} based on plain LSDA calculations.
Therefore, the position of the Fermi level with respect to the band gap
is controlled more by the static rather than by the dynamic
correlations.

In the present work, both static and dynamic correlations are taken into
account simultaneously by combining them as it was done in~\cite{CMK+08}.

\section{Computational details}

The calculations were performed within the relativistic full-potential
Korringa-Kohn-Rostoker Green's function (SPR-KKR) method~\cite{MCP+05}.
The DMFT-solver consists in the relativistic version of the so-called
Spin Polarised T-Matrix Plus Fluctuation Exchange (SPTF) approximation
\cite{KL02,PKL05} using the real-energy axis formalism~\cite{DJK02}.
According to this scheme the local Green's function is obtained by the
corresponding site projection of the full KKR Green's function. The
local Green's function is needed to obtain the bath Green's function
for the Anderson impurity model via the saddle-point equation. The bath
Green's function is used as input for the SPTF DMFT-scheme to calculate
the local self-energy. The latter appears as an additional
energy-dependent potential in the radial Dirac equation which is solved
in order to calculate the new full KKR Green's function. This procedure
is repeated until self-consistency in both the self-energy and the
charge density is achieved. This scheme has been already successfully
applied for the description of magneto-optical~\cite{CME+06} and
magnetic~\cite{CMK+08} properties of 3$d$ transition metals. Also
photoemission intensities~\cite{MCE+05}, including the corresponding
matrix element effects, were obtained in quantitative agreement with
the experiment.

The static double-counting is treated in the form of the so-called
"around atomic limit" (AAL)~\cite{CS94} which is appropriate in the
case of integer orbital occupation numbers and therefore corresponds to
the half-metals. Although, different mechanisms exist that lead to
half-metallicity, in the case of locally correlated half-metals the AAL
tends to increase the spin magnetic moment, which is often
underestimated in plain LSDA calculations~\cite{WFK+05}. The
self-energy within the DMFT is parameterised by the averaged screened
Coulomb interaction $U$ and the Hund exchange interaction $J$. For the
latter the screening is usually not crucial; the value of $J$ can be
calculated directly from LSDA and is approximately equal to 0.9~eV for
all 3$d$ elements. This value has been adopted for all calculations
presented here. Different methods in calculating the screened Coulomb
interaction $U$ for the metallic 3$d$ transition metal systems lead to
$U$-values in the range between 2-3~eV. In our analysis we keep the $U$
parameter fixed at 3~eV which is often used for 3$d$ transition metals.

The random substitutional order of the Mn-Fe compounds is treated
within the so-called Coherent Potential Approximation (CPA). Since the
CPA approach can be formulated in the multiple scattering formalism
\cite{Gyo72,But85} it has a rather straightforward implementation
within the KKR method~\cite{EB96}. As it follows from the experiment
the lattice constant does not change substantially its value within the
whole range of concentrations $x$. The lattice parameter was fixed to
5.658~\AA in the present study.

Concerning the calculation of the ground state properties few
additional details have to be mentioned at this stage. In particular we
stress the importance of the full potential treatment, which often
leads to some improvements in the description of the band gap already
in plain LSDA calculations. On the other hand, one has to increase the
$l$-value for the expansion in spherical harmonics. This typically
improves the calculated value for the band gap and for the magnetic
moments. In our case we used $l_{\rm max}=3$ ($f$~electrons).

The 8~keV valence-band photoemission spectra (VB-XPS) were calculated
using the formalism derived by Ebert and Schwitalla~\cite{ES97}, which
is based on the so-called one-step model of photoemission. In addition,
a quantitative treatment of core-hole effects on the valence states
turned out to be of crucial importance for a correct description of
hard X-ray VB-XPS. Here, the presence of the core-hole was explicitly
accounted for when calculating the initial state. Here, we used the
static so-called $Z+1$ approach~\cite{Alo94}. This means that the
electronic structure has to be calculated self-consistently with a core
state of the excited atom (here $1s$ of Mn) being unoccupied. The
missing charge is added to the valence band.

Another important aspect which one has to consider in calculations of
hard X-ray photoemission is the correct treatment of the X-ray
cross-section as a function of the photon energy. It follows from the
calculations (see \cite{FBO+07}) that the cross-sections for $p$ and
$d$ electrons are nearly the same for kinetic energies of 3-4~keV.
However, at 8~keV their ratios are completely different, in particular
the $d$ cross-section becomes much smaller. In the present work these
effects are taken into account implicitly within the one-step
formalism. On the other hand the uncertainty in the calculation of the
final state increases with increasing photon energy, because of an
increasing number of oscillations in the final state radial wave
functions. Therefore, one needs to account for a much larger number of
points on the radial mesh in order to reproduce these variations
correctly.

In order to account for the experimental resolution, the calculated
spectra were convoluted by a Gaussian with FWHM = 0.025~eV. For a
direct comparison, the background intensity was subtracted from the
experimental data using the Shirley procedure~\cite{Shi72}.

\section{Results and discussion}

\subsection*{Ground-state properties}

\begin{figure}[H]
\begin{center}
  \includegraphics[width=0.5\textwidth]{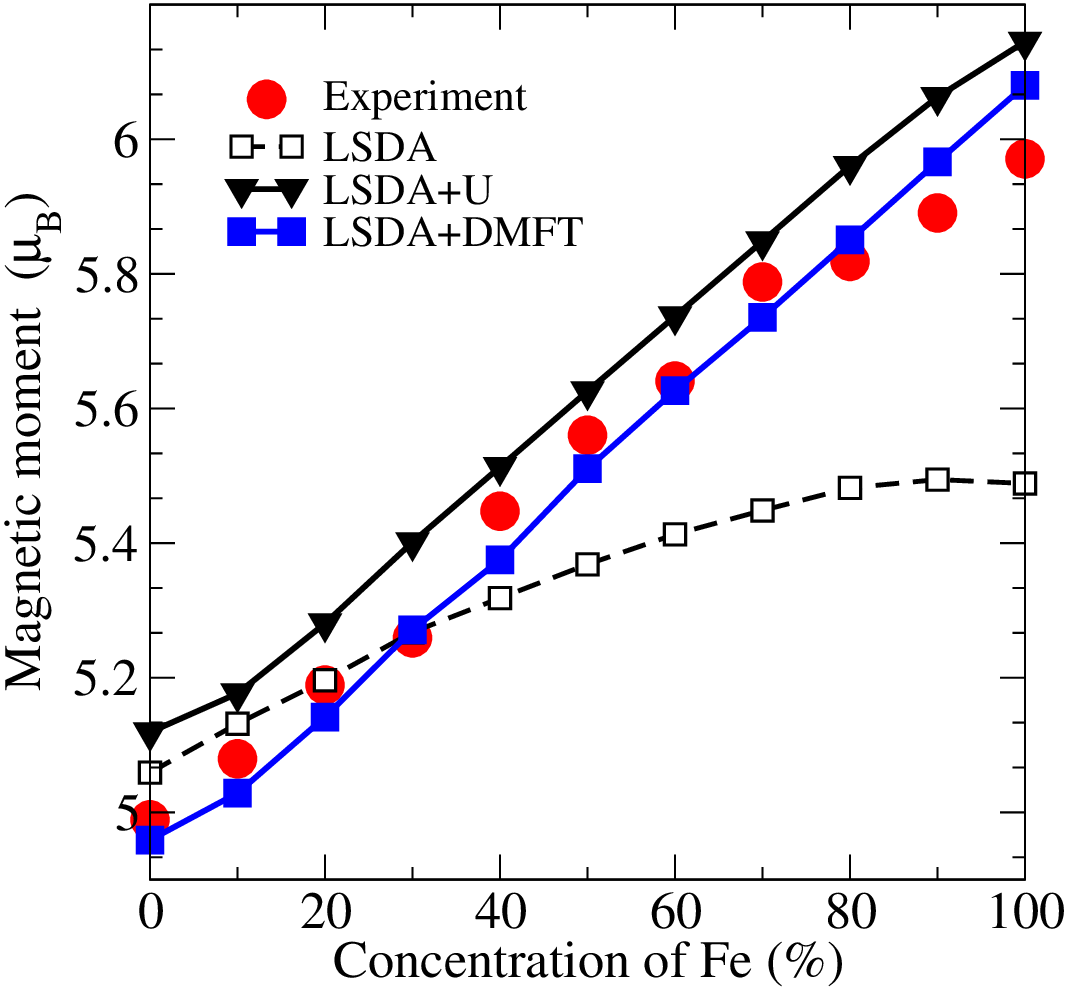}
  \caption{Comparison of  the total magnetic  moments for
Co$_2$Mn$_x$Fe$_{1-x}$Si compounds calculated  within  LSDA (black  dashed
line,  opened squares),  LSDA+U  (black triangles)  and LSDA+DMFT  (blue
squares)   with  the   results  of   the  SQUID   magnetic  
measurements (red circles) \cite{BFKF06}. \label{FIG_MOM}}
\end{center}
\end{figure}

As it was mentioned above, a quantitative treatment of static
correlations is important for a reliable description of the minority
band gap as well as of the magnetic moments in correlated half-metals.
More details can be found in preceding publications~\cite{WFK+05}. Our
calculations show a noticeable increase of the magnetic moments in
analogy to the Slater-Pauling curve within the whole range of
concentrations (see \fref{FIG_MOM}). By construction, the
dynamical part of the local self-energy $\Sigma$ in the vicinity of the
Fermi level describes Fermi liquid behaviour, i.e. ${{\rm
Re}\,\Sigma(\epsilon)\!\sim\!-(\epsilon-\epsilon_{\rm F})}$, ${{\rm
Im}\,\Sigma(\epsilon)\!\sim-(\epsilon-\epsilon_{\rm F})^2}$. Therefore,
we do not expect a substantial change for the integral quantities by
accounting only for dynamical correlations. Indeed, as it follows from
\fref{FIG_MOM}, the magnetic moments calculated with LSDA+DMFT
are only slightly reduced comparing to those obtained from LSDA+U. This
small decrease may be attributed to the spin-flip events induced by the
dynamical fluctuations leading to the so-called non-quasiparticle
states absent in the LSDA treatment~\cite{CSA+08}. At the same time the
effect of static correlations on the magnetic moment is more pronounced
for the Fe-rich limit of the concentration $x$.

\begin{figure}[H]
\begin{center}
  \includegraphics[width=0.5\textwidth]{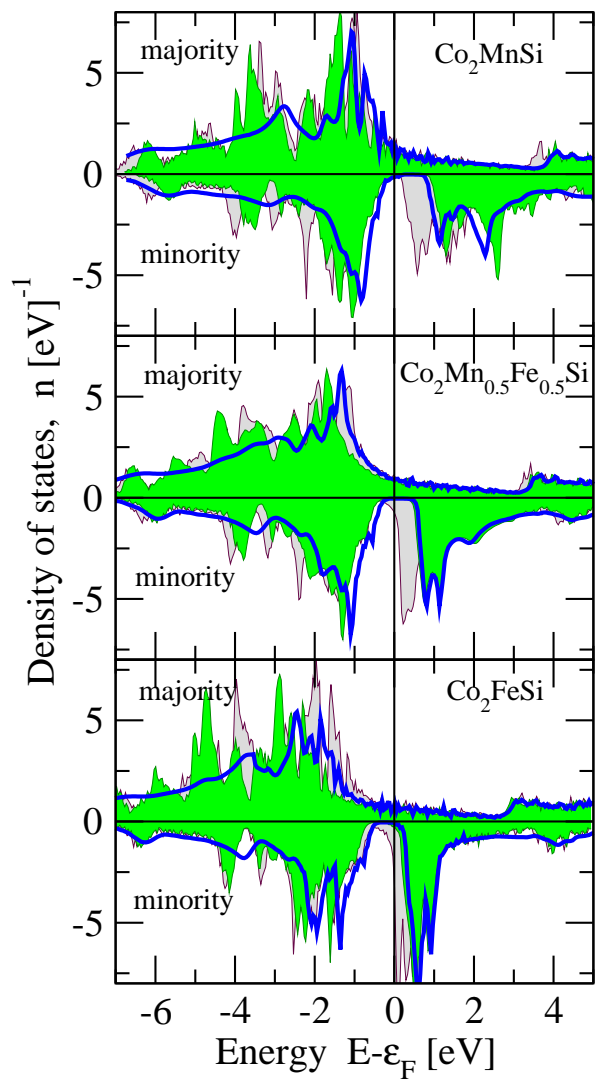}\hspace*{-0.5ex}
  \caption{The  total  spin-resolved  DOS  curves  for
Co$_2$MnSi,  Co$_2$Mn$_{0.5}$Fe$_{0.5}$Si  and
Co$_2$FeSi calculated within  LSDA (light/grey filled area),  LSDA+U
(dark/green filled area) and  LSDA+DMFT  (blue line).\label{FIG_DOS}}
\end{center}
\end{figure}

The dynamical correlations appear to be more significant for the
spectroscopic properties. As it follows from \fref{FIG_DOS}, the
LSDA+U shifts the $d$ states away from the Fermi level creating a band
gap in the minority-spin channel. At the same time adding the 
dynamical correlations tend to narrow the bandwidth by shifting the
states back to the Fermi level, however without influencing the band
gap itself. One also can observe a substantial broadening of the DOS
within the range between $-2$ to $-8$~eV, which is due to the imaginary
part of the energy-dependent self-energy.

It follows from the DOS curves, that the enhancement of the magnetic
moment with the increase of the Fe concentration (see
\fref{FIG_DOS}) is connected to the corresponding shift of the
Fermi energy across the band-gap. In both limiting cases one observes
the non-vanishing minority-spin states at the Fermi level. This
indicates that both Co$_2$MnSi and Co$_2$FeSi may not represent fully
spin polarised ferromagnets in contrast to the compounds with
intermediate Fe concentrations.

\subsection*{Photoemission}

\begin{figure}[H]
\begin{center}
  \includegraphics[width=0.5\textwidth]{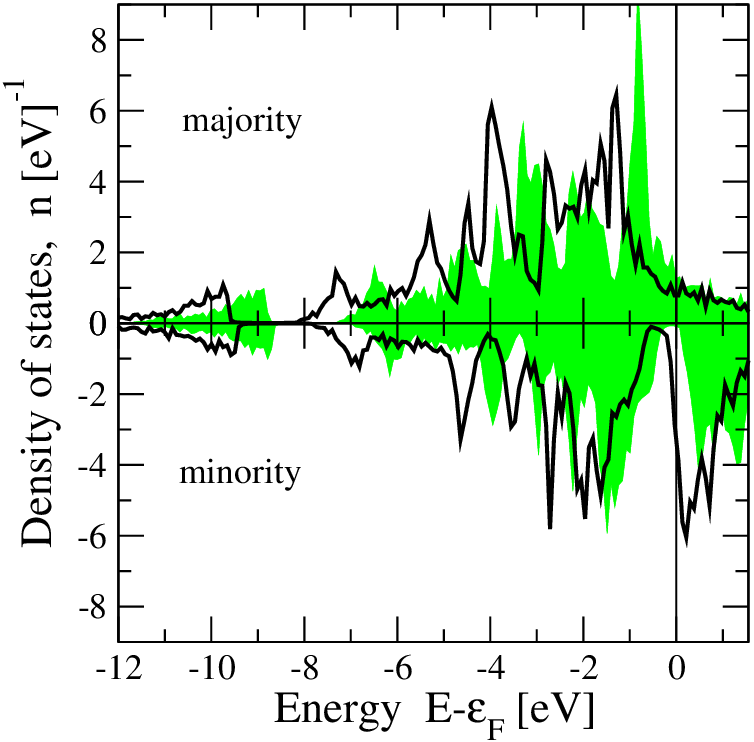}
  \caption{Comparison of  the total  spin-resolved (LSDA)
DOS curves  for Co$_2$MnSi  in case  of the hole  created in  $1s$ core
level of Mn (black line) and without (green filled area).\label{FIG_DOSCORE}}
\end{center}
\end{figure}

It was mentioned above, that an important issue is to account for the
electron interaction with holes created in the photoemission process.
The hole life-time increases when approaching the core energies.
Therefore, the core holes itself have the longest life times and in
consequence the most pronounced influence on the photoemission spectra.
According to the experimentally available core energies the 8~keV
photon beam used in the measurements corresponds to the excitation of
the $1s$ states of Mn. This assumption agrees well with the results
obtained in a preceding investigation~\cite{FBO+07}. Indeed, comparing
the experimental photoemission data to the calculated DOS one notices
that the latter must be scaled on the energy axis by a factor of about
1.1 in case of Co$_2$MnSi. However, in the case of Co$_2$FeSi (with no
manganese) the $-10$~eV $s$-peak is already in good agreement with the
experiment. In the present paper we account for this effect explicitly
by introducing in the self-consistent calculation for Co$_2$MnSi a core
hole in the Mn $1s$ level. The corresponding change in the DOS 
is shown in \fref{FIG_DOSCORE}. Since the account of correlation
effects in the actual photoemission calculations is essentially based
on a ground-state information, the core hole is accounted for just by
simply scaling the final result by a factor of 1.1 times the
concentration of Mn along the energy axis.

\begin{figure}[H]
\begin{center}
   \includegraphics[width=0.5\textwidth,clip]{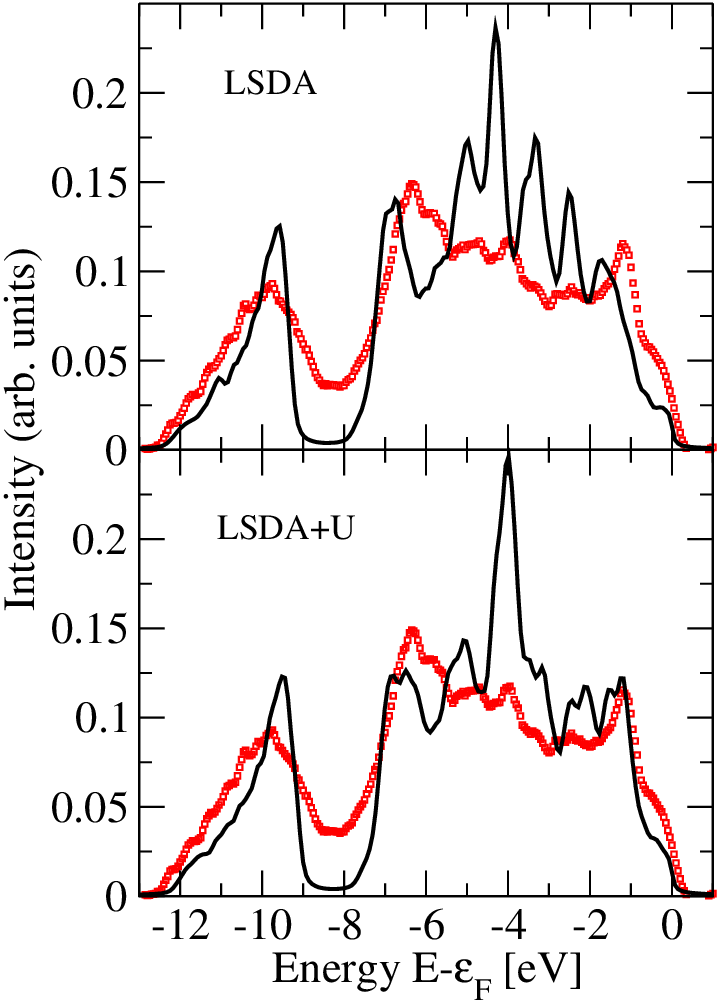}
  \caption{Comparison  of the  theoretical  (solid lines)
8~keV VB-XPS  spectra for Co$_2$MnSi  calculated within plain  LSDA  and
LSDA+U to the experiment (red squares) \cite{FBO+07}. 
 The area under each curve is normalised to unity and
1s core-hole effects of Mn are included. \label{FIG_XPS1}}
\end{center}
\end{figure}

The influence of correlation effects in VB-XPS is illustrated in
Fig.~\ref{FIG_XPS1}. It follows, that the consideration of static
correlations (LSDA+U) substantially improves the plain LSDA result in
the range above $-5$~eV, reproducing the $-1$~eV and $-4$~eV peaks, which are
shifted comparing to plain LSDA by about 0.5~eV to higher binding
energies. Even more, tiny features such as the local minima at $-1.5$~eV,
$-2.5$~eV and the peak splittings are well reproduced. Not only the peak
positions but also the intensities above $-5$~eV were improved in
comparison to the experiment, while only the $-4$~eV peak appears to be
much too intense. Various attempts to treat the final states more
accurately did not changed this situation substantially. Therefore, it
is concluded that the origin of this discrepancy is found in a
band-structure effect caused by the Si $p$ states and by the $d$ states
of Mn and Co. Indeed, the corresponding total DOS spectra (see
\fref{FIG_DOS}) show the most intensive peak at about $-3$~eV
binding energy, which is formed by the corresponding $p$ and $d$
states. As discussed above one has to scale the DOS energy axis by a factor of 1.1 in order
to compare with the VB-XPS. At high photon energies the cross-section
of the $p$ states is larger than for the $d$ states, however the later
ones have larger spectral weight. Therefore, the contributions of $p$
and $d$ states are comparable.

\begin{figure}[H]
\begin{center}
   \includegraphics[width=0.5\textwidth,clip]{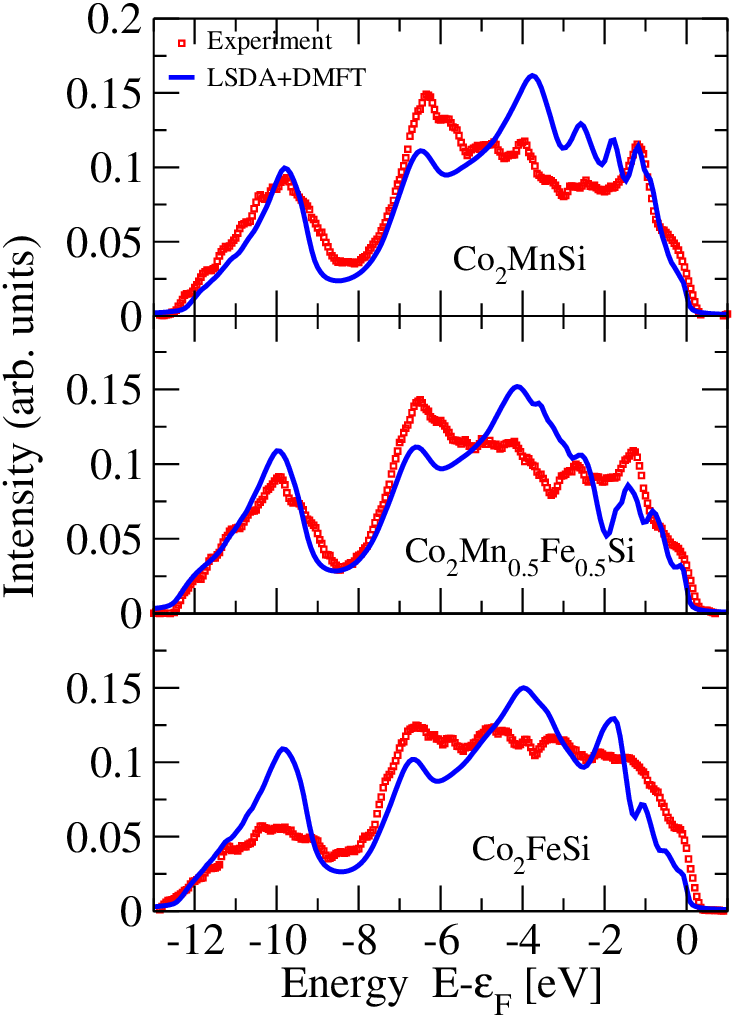}
  \caption{The 8~keV  VB-XPS spectra for  Co$_2$MnSi,  
 Co$_2$Mn$_{0.5}$Fe$_{0.5}$Si and  Co$_2$FeSi
 calculated within  LSDA+DMFT (solid line)  compared to
the experiment  (squares)~\cite{FBO+07}.  The  area under each  curve is
normalised to unity and 1s core-hole effects of Mn are included.
\label{FIG_XPS2}}
\end{center}
\end{figure}

The result of the complete correlation schemes illustrated in the top
panel of \fref{FIG_XPS2}. In conclusion, it follows that
inclusion of DMFT makes some important improvements for the VB-XPS
spectra by correcting the amplitudes and the positions of the $-1$~eV and
$-10$~eV peaks. Also the intermediate area between $-10$ and $-6$~eV binding
energy shows a very close agreement with experiment. The position of
the $-6$~eV peak formed by $p$ states of Si is still not perfectly
reproduced, however it is moved in the proper direction compared to
LSDA+U result. The pronounced spectral feature observable at $-4$~eV is
substantially reduced in intensity, but it is still overestimated by
theory. The position is slightly shifted towards the Fermi level
induced by the real part of the energy-dependent self-energy. The
corresponding shift of the DOS is seen in \fref{FIG_DOS}. The
energy regime ranging from $-8.5$ to $-4.5$~eV binding energy corresponds
to the maximal amplitude of the imaginary component of the self-energy.
From \fref{FIG_XPS2}, it follows that this causes some Lorentzian
over-broadening of the features at $-4.5$~eV for all calculated
concentrations. One the other hand it perfectly describes the so-called
Heusler gap at about $-8.5$~eV. Thus, it is concluded that a non-zero
intensity in the Heusler gap is mostly determined by the electron
ground states existing in this region, rather than by background
effects.

\section{Summary}

From the detailed analysis it follows that Co$_2$Mn$_{1-x}$Fe$_x$Si is
a very pronounced locally correlated material for which the plain LSDA
calculations are insufficient to describe their electronic and magnetic
properties correctly. The values of the magnetic moments and the
details of the minority band gap are mostly determined by static
correlations. In particular, this influence becomes more significant
for the magnetic moments with increasing Fe concentration. A
satisfactory description of this effect can be given on the basis of
the LSDA+U approach within the AAL type of double-counting.

It turned out, that it is not sufficient to explain all spectral
properties accounting for static correlations only. Peak positions and
intensities are significantly improved due to the incorporation of
dynamical correlations. These correlations can be described
quantitatively by the LSDA+DMFT approach.

Obviously, the improvement is not completely perfect and agreement with
experiment gets worse with increasing Fe concentration. One reason may
be found in the perturbative nature of the used DMFT-solver that
affects the ground-state description. However, this is difficult to
verify, since more accurate solvers are much more time-consuming and
restricted to only very small systems. In principle, the DMFT itself
becomes exact only in the limit of infinite coordination
numbers~\cite{KSH+06}. This implies that non-local correlations may
also play a role~\cite{SHR+05}. The uncertainty of the double-counting
and values of $U$ and $J$ has to be considered, too. Another source of
errors is the insufficient description of the excited properties, in
particular the uncertainty in dealing with the final states. On the
other hand, a very important excitation effect, namely the influence of
the core holes was recognised, quantified and taken into account in the
present work.

Finally, it is emphasised that the LSDA+DMFT scheme used in this work
has significantly improved the description of the magnetism and VB-XPS
in the presented series of Heusler compounds with $L2_1$ structure.
This result is in accordance with former studies on 3$d$ transition
metals.

\ack

The authors like to thank K.~Kobayashi, E.~Ikenaga, J.-J.~Kim, and S.~Ueda for
support with the photoemission experiments at Spring-8 (Japan).
The HAXPES measurements were performed under
the approval of NIMS Beamline Station (Proposal Nos. 2007A4903,
2007B4907) and at BL47XU of SPring-8 (Japan) under the approval of
JASRI (Proposal Nos. 2006A1476,  2008A0017). 
The authors gratefully acknowledge financial
support  by the Deutsche Forschungsgemeinschaft DFG (Research Unit FOR~559, Project P~7).

\newpage

\section*{References}

\bibliographystyle{unsrt}

\begin{thebibliography}{10}

\bibitem{KWS83}
J~K\"{u}bler, A~R Williams, and C~B Sommers.
\newblock {\em Phys.~Rev.~B}, 28:1745, 1983.

\bibitem{FFB07}
C~Felser, G~H Fecher, and B~Balke.
\newblock {\em Angew. Chem. Int. Ed.}, 46:668, 2007.

\bibitem{FSIA90}
S~Fuji, S~Sugimura, S~Ishida, and S~Asano.
\newblock {\em J.~Phys.~Cond.~Matt.}, 2:8583, 1990.

\bibitem{BNWZ00}
P~J Brown, K-U Normann, P~J Webster, and K~R~A Ziebeck.
\newblock {\em J.~Phys.~Cond.~Matt.}, 12:1827, 2000.

\bibitem{FKW+06}
G~H Fecher, H~C Kandpal, S~Wurmehl, C~Felser, and G~Sch{\"o}nhense.
\newblock {\em J. Appl. Phys.}, 99:08J106, 2006.

\bibitem{KFF07b}
J~K{\"u}bler, G~H Fecher, and C~Felser.
\newblock {\em Phys. Rev. B}, 76:024414, 2007.

\bibitem{WFK+05}
S~Wurmehl, G~H Fecher, H~C Kandpal, V~Ksenofontov, C~Felser, and H-J Lin.
\newblock {\em Phys.~Rev.~B}, 72:184434, 2005.

\bibitem{WFK+06}
S~Wurmehl, G~H Fecher, H~C Kandpal, V~Ksenofontov, C~Felser, and H-J Lin.
\newblock {\em Appl.~Phys.~Lett.}, 88:032503, 2006.

\bibitem{KFF06}
H~C Kandpal, G~H Fecher, and C~Felser.
\newblock {\em Phys.~Rev.~B}, 73:094422, 2006.

\bibitem{BFK+06}
B~Balke, G~H Fecher, H~C Kandpal, C~Felser, K~Kobayashi, E~Ikenaga, J-J Kim,
  and S~Ueda.
\newblock {\em Phys.~Rev.~B}, 74:104405, 2006.

\bibitem{Kub00}
J~K{\"u}bler.
\newblock {\em Theory of Itinerant Electron Magnetism}.
\newblock Clarendon Press, Oxford, 2000.

\bibitem{GDP02}
I~Galanakis, P~H Dederichs, and N~Papanikolaou.
\newblock {\em Phys. Rev. B}, 66:174429, 2002.

\bibitem{KFFS06}
H~C Kandpal, G~H Fecher, C~Felser, and G~Sch\"onhense.
\newblock {\em Phys.~Rev.~B}, 73:094422, 2006.

\bibitem{FBO+07}
G~H Fecher, B~Balke, S~Ouardi, C~Felser, G~Sch\"onhense, E~Ikenaga, J~Kim,
  S~Ueda, and K~Kobayashi.
\newblock {\em J.~Phys.~D: Appl.~Phys.}, 40:1576, 2007.

\bibitem{AZA91}
V~I Anisimov, J~Zaanen, and O~K Andersen.
\newblock {\em Phys.~Rev.~B}, 44:943, 1991.

\bibitem{CS94}
M~T Czyzyk and G~A Sawatzky.
\newblock {\em Phys.~Rev.~B}, 49:14211, 1994.

\bibitem{KSH+06}
G~Kotliar, S~Y Savrasov, K~Haule, V~S Oudovenko, O~Parcolett, and C~A
  Marianetti.
\newblock {\em Rev.~Mod.~Phys.}, 78:865, 2006.

\bibitem{LKK01}
A~I Lichtenstein, M~I Katsnelson, and G~Kotliar.
\newblock {\em Phys.~Rev.~Lett.}, 87:067205, 2001.

\bibitem{GDK+06}
A~Grechnev, I~Di Marco, M~I Katsnelson, A~I Lichtenstein, J~Wills, and
  O~Eriksson.
\newblock {\em cond-mat}, page 0610621, 2006.

\bibitem{PCE03}
A~Perlov, S~Chadov, and H~Ebert.
\newblock {\em Phys.~Rev.~B}, 68:245112, 2003.

\bibitem{MEN+05}
J~Min\'{a}r, H~Ebert, C~de~Nada\"{i}, N~B Brookes, F~Venturini, G~Ghiringhelli,
  L~Chioncel, A~I Lichtenstein, and M~I Katsnelson.
\newblock {\em Phys.~Rev.~Lett.}, 95:166401, 2005.

\bibitem{BME+06}
J~Braun, J~Min\'{a}r, H~Ebert, M~I Katsnelson, and A~I Lichtenstein.
\newblock {\em Phys.~Rev.~Lett.}, 97:227601, 2006.

\bibitem{CMK+08}
S~Chadov, J~Min\'{a}r, M~I Katsnelson, H~Ebert, D~K\"{o}dderitzsch, and A~I
  Lichtenstein.
\newblock {\em Europhys.~Lett.}, 82:37001, 2008.

\bibitem{KL02}
M~I Katsnelson and A~Lichtenstein.
\newblock {\em Europ.~Phys.~J.~B}, 30:9, 2002.

\bibitem{CSA+08}
L~Chioncel, Y~Sakuraba, E~Arrigoni, M~I Katsnelson, M~Oogane, Y~Ando,
  T~Miyazaki, E~Burzoa, and A~I Lichtenstein.
\newblock {\em Phys.~Rev.~Lett.}, 100:086402, 2008.

\bibitem{IKL07}
V~Yu Irkin, M~I Katsnelson, and A~I Lichtenstein.
\newblock {\em J.~Phys.~Cond.~Matt.}, 19:315201, 2007.

\bibitem{IFKA95}
S~Ishida, S~Fuji, S~Kashiwagi, and S~Asano.
\newblock {\em J.~Phys.~Soc.~Jap.}, 64:2152, 1995.

\bibitem{MCP+05}
J~Min\'{a}r, L~Chioncel, A~Perlov, H~Ebert, M~I Katsnelson, and A~I
  Lichtenstein.
\newblock {\em Phys.~Rev.~B}, 72:45125, 2005.

\bibitem{PKL05}
L~V Pourovskii, M~I Katsnelson, and A~I Lichtenstein.
\newblock {\em Phys.~Rev.~B}, 72:115106, 2005.

\bibitem{DJK02}
V~Drchal, V~Jani\u{s}, and J~Kudrnovsky.
\newblock {\em Physica~B}, 312-313:519, 2002.

\bibitem{CME+06}
S~Chadov, J~Min\'{a}r, H~Ebert, A~Perlov, L~Chioncel, M~I Katsnelson, and A~I
  Lichtenstein.
\newblock {\em Phys.~Rev.~B}, 74:140411(R), 2006.

\bibitem{MCE+05}
J~Min\'ar, S~Chadov, H~Ebert, L~Chioncel, A~Lichtenstein, C~de~Nada\"{i}, and
  N~B Brookes.
\newblock {\em Nuc. Inst. and Met. in Phys. Res. A}, 574:151, 2005.

\bibitem{Gyo72}
B~Gyorffy.
\newblock {\em Phys.~Rev.~B}, 5:2382, 1972.

\bibitem{But85}
W~H Butler.
\newblock {\em Phys.~Rev.~B}, 31:3260, 1985.

\bibitem{EB96}
H~Ebert and M~Battocletti.
\newblock {\em Solid~State~Comm.}, 98:785, 1996.

\bibitem{ES97}
H~Ebert and J~Schwitalla.
\newblock {\em Phys.~Rev.~B}, 55:3100, 1997.

\bibitem{Alo94}
M~Alouani.
\newblock {\em Phys.~Rev.~B}, 49:16038, 1994.

\bibitem{Shi72}
D~A Shirley.
\newblock {\em Phys.~Rev.~B}, 5:4709, 1972.

\bibitem{BFKF06}
B~Balke, G~H Fecher, H~C Kandpal, and C~Felser.
\newblock {\em Phys.~Rev.~B}, 74:104405, 2006.

\bibitem{SHR+05}
J~Sch\"{a}fer, M~Hoinkis, E~Rotenberg, P~Blaha, and R~Claessen.
\newblock {\em Phys.~Rev.~B}, 72:155115, 2005.

\end{thebibliography}

\end{document}